\documentclass[fleqn,usenatbib]{mnras}

\usepackage{newtxtext,newtxmath}
\usepackage[T1]{fontenc}

\usepackage{graphicx}	% Including figure files
\usepackage{amsmath}	% Advanced maths commands
\usepackage[normalem]{ulem}
\usepackage{bm}
\title[$f\left(R\right)$ gravity versus the Hubble tension]{$f\left(R\right)$ gravity in the Jordan frame as a paradigm for the Hubble tension}

\author[T. Schiavone, G. Montani \& F. Bombacigno]{
Tiziano Schiavone,$^{1,2,3}$\thanks{E-mail: tschiavone@fc.ul.pt}
Giovanni Montani,$^{4,5}$\thanks{E-mail: giovanni.montani@enea.it}
Flavio Bombacigno,$^{6}$\thanks{E-mail: flavio2.bombacigno@uv.es}
\\
$^{1}$Department of Physics ``E. Fermi", University of Pisa, Polo Fibonacci, Largo Bruno Pontecorvo 3, I-56127 Pisa, Italy\\
$^{2}$INFN, Istituto Nazionale di Fisica Nucleare, Sezione di Pisa, Polo Fibonacci, Largo Bruno Pontecorvo 3, I-56127~Pisa, Italy\\
$^{3}$Instituto de Astrofisíca e Ciências do Espaço, Faculdade de Ciências da Universidade de Lisboa, Edificio C8, Campo Grande, P-1740-016, Lisbon, Portugal\\
$^{4}$ENEA, Fusion and Nuclear Safety Department, C.R. Frascati, Via Enrico Fermi 45,  Frascati, I-00044 Rome, Italy\\
$^{5}$Physics Department, ``Sapienza" University of Rome, Piazzale Aldo Moro 5, I-00185 Rome, Italy\\
$^{6}$Departament de F\'{i}sica Teòrica and IFIC, Universitat de València, Carrer del Doctor Moliner 50, E-46100, Burjassot, Spain\\
}

\date{Accepted 2023 March 22. Received 2023 March 21; in original form 2022 December 2}

\pubyear{2023}
\volume{522}
\pagerange{L72--L77}

\begin{document}
\label{firstpage}
\maketitle

% It should be a single paragraph not more than 250 words (200 words for Letters).
% No references should appear in the abstract.
\begin{abstract}
We analyse the $f(R)$ gravity in the so-called Jordan frame, as implemented to the isotropic Universe dynamics. The goal of the present study is to show that, according to recent data analyses of the supernovae Ia Pantheon sample, it is possible to account for an effective redshift dependence of the Hubble constant. This is achieved via the dynamics of a non-minimally coupled scalar field, as it emerges in the $f(R)$ gravity. We face the question both from an analytical and purely numerical point of view, following the same technical paradigm. We arrive to establish that the expected decay of the Hubble constant with the redshift $z$ is ensured by a form of the scalar field potential, which remains essentially constant for $z\lesssim0.3$, independently if this request is made \textit{a priori}, as in the analytical approach, or obtained \textit{a posteriori}, when the numerical procedure is addressed. Thus, we demonstrate that an $f(R)$ dark energy model is able to account for an apparent variation of the Hubble constant due to the rescaling of the Einstein constant by the $f(R)$ scalar mode.
\end{abstract}


%%%%%%%%%%%%%%%%%%%%%%%%%%%%%%%%%%%%%%%%%%%%%%%%%%

%%%%%%%%%%%%%%%%% BODY OF PAPER %%%%%%%%%%%%%%%%%%

\section{Introduction}\label{sec:intro}

The measurement of the Hubble constant $H_0$ has been one of the most challenging effort of large-scale observations of the present Universe since the very beginning of cosmological studies. For many decades, the value of $H_0$ has been determined with a very low degree of precision. However, since the beginning of the new century, the emergence of the so-called 'precision cosmology' allowed accurate measurements of $H_0$, and the possibility to test an increasingly large set of cosmological parameters. Nowadays, a large number of different and independent measurements of $H_0$ are available \citep{DiValentino:2021izs} and also other crucial cosmological indicators can be found, like the position of peaks in the cosmic microwave background (CMB) data thanks to the \textit{Planck} satellite \citep{Planck:2018vyg}. 

However, the end of the previous century was characterized by a big surprise, coming from low-redshift observations of Type Ia supernovae (SNe Ia) recovered as standard candles: the present Universe is accelerating \citep{SupernovaSearchTeam:1998fmf,SupernovaCosmologyProject:1998vns}. This experimental issue opened a new era on the understanding of the present Universe physics, since either a dark energy component [identified in a cosmological constant term in the so-called lambda cold dark matter ($\Lambda$CDM) model \citep{Weinberg:2008zzc}], or a modified gravity theory must be postulated to represent the emerging acceleration. 
\\ \indent In this very puzzling panorama, in recent years an additional non-trivial observational evidence came out, called the Hubble tension \citep{DiValentino:2021izs}, i.e. a discrepancy in 4.9~$\sigma$ between the determination of $H_0$ via the CMB data ($H_{0}^{\text{CMB}}=67.4\pm0.5\,\textrm{km s}^{-1}\,\textrm{Mpc}^{-1}$) \citep{Planck:2018vyg} and the local one coming out by using low-redshift cosmological testers, like the Cepheid-SN Ia sample ($H_{0}^{\text{loc}}=73.04\pm1.04\,\textrm{km s}^{-1}\,\textrm{Mpc}^{-1}$) \citep{pantheon+}.
\\ \noindent Such a discrepancy appears hard to be straightforwardly interpreted, and different values of $H_0$ can be essentially traced back to two causes: either the astrophysical characterization of the tester is inadequate (for instance, because their calibration or redshift evolution are not properly fixed) or new physics, in addition to the dark energy Universe component, must be considered \citep{Vagnozzi:2019ezj}. The analysis pursued by the SNe Ia community, mainly represented by the data of the Pantheon \citep{scolnic-pantheon,https://doi.org/10.17909/t95q4x} and Pantheon$+$ \citep{pantheon+} samples, seem to exclude the existence of a redshift evolution of these objects and show a reliable control of all the main sources of errors. However, some recent studies report a possible redshift dependence of the marginalized absolute magnitude of SNe Ia \citep{kazantzidis}, or the Hubble constant itself \citep{Krishnan:2020obg,dainottiApJ-H0(z),Krishnan:2020vaf,colgain-bin,Colgain:2022tql,DainottiGalaxies-H0(z),Jia:2022ycc,Krishnan:2022,Schiavone:2022shz,2023arXiv230110572D}. 
\\ \indent In particular, the two analyses by \cite{dainottiApJ-H0(z),DainottiGalaxies-H0(z)}, based on a binned distribution of the SNe Ia with equipopulated bins, have outlined a variation of $H_0$ as $(1+z)^{-\alpha}$ within 2~$\sigma$ confidence level, with $\alpha\sim 10^{-2}$. The evolution of variables in astrophysics is a well-studied subject, and it has been discussed already in the realm of gamma-ray bursts (GRBs) \citep{Dainotti:2020azn}. Furthermore, in \cite{dainottiApJ-H0(z)}, it has been stressed that the extrapolation of the behaviour of $H_{0}^{\text{fit}}(z)$ up to the CMB redshift seems to naturally account for the Hubble tension feature, since a larger value of the Hubble constant today would slowly decrease to higher redshifts. Other recent studies analysed and discussed the assumption of Gaussian likelihoods in evaluating cosmic distances to obtain cosmological constraints by using SNe \citep{Dainotti:2023ebr}, GRBs, quasars, and baryonic acoustic oscillations \citep{Bargiacchi:2023jse}.
\\ \indent Among several proposals for alleviating the $H_0$ tension \citep{DiValentino:2021izs}, an interesting possibility is provided by modified gravity theories. In particular, in \cite{dainottiApJ-H0(z)}, it was discussed that the observed dependence of an effective Hubble constant $H_{0}^{\text{eff}}(z)$, predicted by the binning analysis of the Pantheon sample, can be interpreted as a variation of the Einstein constant, naturally achieved for example by $f(R)$ gravity in the Jordan frame \citep{Olmo:2005hc,Olmo:2005zr,Nojiri-Odintsov2007,Nojiri:2010wj-unified,Olmo:2006eh,Sotiriou-Faraoni2010,Faraoni:2010pgm,Nojiri:2017ncd-nutshell}. These theories are endowed with an additional scalar degree of freedom non-minimally coupled to the metric [see \cite{Moretti:2019yhs} for a gauge-invariant analysis], which can be used in principle for addressing unsolved problems in the $\Lambda$CDM model, such as the $H_0$ tension \citep{DiValentino:2021izs,Odintsov:2020qzd,Nojiri:2022ski}. However, in \cite{DainottiGalaxies-H0(z)} it was shown how one of the most reliable $f(R)$ models, the Hu--Sawicki proposal \citep{Hu:2007nk}, is inappropriate to reproduce the desired effect. This negative result suggested the necessity to consider an alternative dark energy $f(R)$ model, able to account both for the Universe acceleration and a variable $H_{0}^{\text{eff}}(z)$ parameter. 

The present letter is dedicated to the formulation of a model satisfying such requirements, following the prescriptions in \cite{dainottiApJ-H0(z),DainottiGalaxies-H0(z)} of a decreasing trend for $H_{0}^{\text{eff}}(z)$. Then, under few assumptions, we derive the profile of the potential term for the scalar field, which in turn allows us to reconstruct the underlying $f(R)$ model.

The analysis is divided into two parts: the first one is characterized by an analytical approach, while the second one relies on a pure numerical study. The analytical formulation starts with the hypothesis that the potential can be satisfactorily described by a dynamical deviation from a flat region (encoding the dark energy contribute for low redshifts). Unlike the analysis in \cite{dainottiApJ-H0(z),DainottiGalaxies-H0(z)}, we do not fix a priori the form of $\phi(z)$, but we obtain it from the dynamics in the Jordan frame. In the numerical analysis, we assume again the evolution of $H_{0}^{\text{eff}}(z)$, but we relax the request of a constant potential term in a given region. It is remarkable that both the analytical and the numerical formulations are consistent and predict a flat potential profile in a region $0<z\lesssim 0.3$. 

An important consistency check is provided by the determination of the $f(R)$ form in the limit of the low redshift. We get three contributions: a cosmological constant, a linear contribution in the Ricci scalar $R$, and eventually a quadratic correction as in the $R^2$-gravity theory \citep{STAROBINSKYquadratic}. It is worth emphasizing that this modified theory reduces to the standard $\Lambda$CDM model if the function $H_{0}^{\text{eff}}(z)$ is frozen to a constant value and $df/dR \equiv 1$ today.

This work is organized as follows: in Sect.~\ref{sec:f(R)inJF} we briefly introduce the $f(R)$ modified gravity in the Jordan frame within the framework of a homogeneous and isotropic Universe; in Sect.~\ref{sec:approx} we derive the scalar field potential, inferred from a running Hubble constant with the redshift; in Sect.~\ref{sec:numerical} we provide our numerical solutions; in Sect.~\ref{sec:lowz} we obtain the functional form of $f(R)$ in the low-redshift limit; in Sect.~\ref{sec:conclusions} we summarize our key findings.  

The metric signature adopted here is $\left(-, +, +, +\right)$, and
the speed of light is $c = 1$. The Newton constant is denoted with $G$, while the Einstein constant is defined as $\chi\equiv8\,\pi\,G$.

\section{f(R) gravity in the Jordan frame for a homogeneous and isotropic Universe}\label{sec:f(R)inJF}

In metric $f(R)$ modified gravity an extra scalar degree of freedom with respect to General Relativity (GR) occurs, by virtue of a Lagrangian density where the Ricci scalar $R$ is replaced by a generic function $f(R)$. This is manifested in the so-called Jordan frame \citep{Olmo:2005hc,Olmo:2005zr}, where the original $f(R)$ theory is restated in the scalar--tensor form:
\begin{equation}
S_{J}=\frac{1}{2\,\chi}\,\int d^{4}x\,\sqrt{-g}\,\left[\phi\,R-V\left(\phi\right)\right]+S_{M}\left(g_{\mu\nu},\psi\right)\,,
\label{eq: azione jordan frame}
\end{equation}
where $g$ is the determinant of the metric tensor, $S_M$ is the action for matter fields $\psi$. Note that in the Jordan frame the additional degree of freedom is defined as $\phi=f^{\prime}\left(R\right)=df/dR$, and it is controlled by the scalar field potential $V\left(\phi\right)=\phi\,R\left(\phi\right)-f\left(R\left(\phi\right)\right)$.

Considering a flat Friedmann--Lemaitre--Robertson--Walker (FLRW) metric \citep{Weinberg:2008zzc}, we can derive the generalized Friedmann equation, the acceleration equation and the scalar field equation as follows\footnote{We remark that the variation of the action with respect to the scalar field $\phi$ actually results in the equation $R=\frac{d V}{d \phi}$. It is this last expression, combined with the trace of the equation for the metric $g_{\mu\nu}$, which results in Eq.~\eqref{eq:scalar-fieldFLRW}.}:
\begin{subequations}
\begin{align}
& H^{2} =\frac{\chi\,\rho}{3\,\phi}+\frac{V\left(\phi\right)}{6\,\phi}-H\,\frac{\dot{\phi}}{\phi}\label{eq:generalized-Friedmann}\\
& \frac{\ddot{a}}{a} =-\frac{\chi}{6\,\phi}\,\left(\rho+3P\right)+\frac{V\left(\phi\right)}{6\,\phi}-\frac{H}{2}\,\frac{\dot{\phi}}{\phi}-\frac{1}{2}\,\frac{\ddot{\phi}}{\phi}\label{eq:generalized-acc}\\
& 3 \ddot{\phi}-2\,V\left(\phi\right)+\phi\,\frac{dV}{d\phi}+9\,H\,\dot{\phi}=\chi\,\left(\rho-3P\right),\label{eq:scalar-fieldFLRW}
\end{align}
\end{subequations}
where $\dot{}=d/dt$, being $t$ the cosmic time in the synchronous gauge, $H(t)$ the Hubble parameter, $\rho\left(t\right)$ and $P\left(t\right)$ the energy density and pressure of the cosmological fluid, respectively. 

Moreover, the divergenceless of the stress-energy tensor for a perfect fluid gives
\begin{equation}
\dot{\rho}+3\,H\,\left(\rho+P\right)=0\,.
\label{eq:continuity}
\end{equation}
To solve  Eqs.~\eqref{eq:generalized-Friedmann} and ~\eqref{eq:scalar-fieldFLRW}, which are the two independent equations, one must also specify the equation of state, that for a barotropic fluid is just $P\left(\rho\right)=w\,\rho$, where $w=0$ and $w=-1$ hold for matter and cosmological constant components, respectively.

Now, as it can be observed in Eq.~\eqref{eq:generalized-Friedmann}, an effective Einstein constant $\chi / \phi$ emerges, whose value ultimately depends on the dynamics of the scalar field.

Cosmological models based on $f(R)$ modified theories have been employed to predict deviations from the $\Lambda$CDM model and can mimick cosmic acceleration in late times, without a true cosmological constant term in the action \citep{Hu:2007nk,Starobinsky:2007hu,Tsujikawa:2007xu}.

%---------------------------------------------------------------------------------------------------------------------
\section{Analytic solution for the scalar field potential}\label{sec:approx}

In order to build the profile of the scalar field potential $V\left(\phi\right)$, we assume the presence of an effective Hubble constant $H_{0}^{\text{eff}}(z)$ evolving with the redshift. The results of the analysis performed in \cite{dainottiApJ-H0(z),DainottiGalaxies-H0(z)} suggested the parametrization
\begin{equation}
    H_{0}^{\text{fit}}\left(z\right)=\frac{\Tilde{H}_0}{\left(1+z\right)^{\alpha}}\,,
    \label{eq:H0(z)}
\end{equation}
where the constants $\alpha$ and $\Tilde{H}_{0}$ are the fitting parameters of the analysis. Note that a decreasing trend with the redshift may address the Hubble tension, since the extrapolation of the fitting function $H_{0}^{\text{fit}}(z)$ from $z=0$ to the recombination redshift $z=1100$ might successfully match $H_{0}^{\text{loc}}$ and $H_{0}^{\text{CMB}}$ \citep{dainottiApJ-H0(z),DainottiGalaxies-H0(z)}. Then, we build the Hubble function $H(z)$ as:
\begin{equation}
    H\left(z\right)=H_{0}^{\text{eff}}(z)\,\sqrt{\Omega_{m0}\left(1+z\right)^3+1-\Omega_{m0}}\,,
    \label{eq:Hubble-parameter(z)}
\end{equation}
where $\Omega_{m0}$ is the cosmological density parameter for the matter component. Moreover, we focus on a cosmological dust in the late Universe (matter with $P=0$), i.e. we neglect relativistic components, and we set $\rho=\rho_{0}\left(1+z\right)^3$ by solving the continuity equation ~\eqref{eq:continuity} with $\rho_0$ the present-day matter density. 

To reconstruct the evolution of $H_{0}^{\text{eff}}(z)$, we compare the phenomenological Hubble function $H(z)$ given by Eq.~\eqref{eq:Hubble-parameter(z)} and the generalized Friedmann equation \eqref{eq:generalized-Friedmann}. Considering that in a homogeneous Universe we have $\phi=\phi\left(z\right)$, Eq.~\eqref{eq:generalized-Friedmann} rewrites
\begin{equation}
    H^{2} =\frac{1}{\phi-\left(1+z\right)\,\phi^{\prime}}\frac{\chi}{3}\left(\rho+\frac{V\left(\phi\right)}{2\,\chi}\right)\,,
    \label{eq:generalized-Friedmann2-with-factor}
\end{equation}
where $\phi^\prime\equiv d\phi/dz$. We used the definition of redshift $a_{0}/a=1+z$ with the standard assumption that the scale factor today is $a_{0}=1$, and also the fact that $dz/dt=-\left(1+z\right)\,H\left(z\right)$.

We define the potential as 
\begin{equation}
    V\left(\phi\right)\equiv 2\chi\rho_{\Lambda}+g\left(\phi\right)\,,
    \label{eq:split-potential}
\end{equation}
where $\rho_{\Lambda}$ is the present value of the Universe dark energy density, and $g\left(\phi\right)$ is the deviation from a cosmological constant scenario. To rewrite Eq.~\eqref{eq:generalized-Friedmann2-with-factor} in a form similar to Eq.~\eqref{eq:Hubble-parameter(z)} and discuss the $\Lambda$CDM limit, we assume the existence of a region in which $g\left(\phi\right)\ll2\chi\rho_{\Lambda}$ for $0<z\lesssim z^*$, where $z^{*}\sim 0.3$ is the redshift of matter-dark energy equivalence. Hence, considering only the constant term in $V\left(\phi\right)$, we rewrite Eq.~\eqref{eq:generalized-Friedmann2-with-factor} as
\begin{equation}
    H^{2} =\frac{H_0^2}{\phi-\left(1+z\right)\,\phi^{\prime}}\left[\Omega_{m0}\left(1+z\right)^3+1-\Omega_{m0}\right]\,,
    \label{eq:generalized-Friedmann3-similar-form}
\end{equation}
where we used the definitions of the critical energy density of the Universe today $\rho_{c0}=3H_{0}^{2}/\chi$ and also of the cosmological density parameters $\Omega_{m0}=\rho_{0}/\rho_{c0}$ and $\Omega_{\Lambda0}=\rho_{\Lambda}/\rho_{c0}=1-\Omega_{m0}$ (flat Universe). Note also that $z^*$ is defined such that $\Omega_{m0}\left(1+z^{*}\right)^{3}=\Omega_{\Lambda0}$. From Eq.~\eqref{eq:generalized-Friedmann3-similar-form} one can recognize the usual terms in the Friedmann equation in the $\Lambda$CDM scenario, up to a factor related to the scalar field $\phi$. Indeed, comparing Eqs.~\eqref{eq:Hubble-parameter(z)} and \eqref{eq:generalized-Friedmann3-similar-form}, we can define the effective Hubble constant
\begin{equation}
    H_{0}^{\text{eff}}(z)=\frac{H_0}{\sqrt{\phi-\left(1+z\right)\,\phi^{\prime}}}\,.
    \label{eq:effective-H0}
\end{equation}

Let us now take into account the scalar field equation \eqref{eq:scalar-fieldFLRW}. Using the relation $\phi=\phi\left(z\right)$ and the approximation for the scalar field potential $V\left[\phi\left(z\right)\right]\approx 2\chi\rho_{\Lambda}$, we obtain:
\begin{align}
& 3\,H^{2}\left(1+z\right)\,\left[\left(1+z\right)\,\phi^{\prime\prime}-\phi^{\prime}\right]-3\,\left(1+z\right)\,\frac{\ddot{a}}{a}\,\phi^{\prime}+\phi\,\frac{dV}{d\phi}=\nonumber\\
& = \chi\,\left(\rho+4\rho_{\Lambda}\right)\,.\label{eq:scalar-field-z}
\end{align}
It should be emphasized that we do not neglect the term $dV/d\phi=dg/d\phi$, since we want to check \textit{a posteriori} the viability of the approximation for the scalar field potential at low redshifts.

Furthermore, by substituting the term $\ddot{\phi}$ from Eq.~\eqref{eq:scalar-fieldFLRW} in Eq.~\eqref{eq:generalized-acc}, we have
\begin{equation}
\frac{\ddot{a}}{a} =-\frac{\chi}{3\,\phi}\,\left(\rho+\rho_{\Lambda}\right)+\frac{1}{6}\frac{dV}{d\phi}-H^{2}\,\left(1+z\right)\frac{\phi^{\prime}}{\phi}\,.\label{eq:generalized-acc-z}
\end{equation}

Then, we combine Eqs.~\eqref{eq:Hubble-parameter(z)}, \eqref{eq:scalar-field-z}, and \eqref{eq:generalized-acc-z}, and we obtain
\begin{align}
& \frac{d\Tilde{V}}{dz}=\frac{1}{\frac{\phi}{\phi^{\prime}}-\frac{1+z}{2}}\left\{3\left[\left(1+z\right)^{3}+\frac{1-\Omega_{m0}}{\Omega_{m0}}\right]\left[1-\left(1+z\right)\frac{\phi^{\prime}}{\phi}+\right.\right.\nonumber\\
& \left.\left.-\left(1+z\right)\frac{H_{0}^{\text{eff\,2}}(z)}{H_{0}^{2}}\left(\left(1+z\right)\left(\phi^{\prime\prime}+\frac{\phi^{\prime2}}{\phi}\right)-\phi^{\prime}\right)\right]+9\frac{1-\Omega_{m0}}{\Omega_{m0}}\right\},\label{eq:dV/dz}
\end{align}
where we rescaled the potential as a dimensionless quantity $\Tilde{V}\equiv V/m^2$ with the constant $m^{2}\equiv \chi\rho_{0}/3=H_{0}^{2}\Omega_{m0}$.
To reproduce a decreasing trend for $H_{0}^{\text{eff}}(z)$ similar to $H_{0}^{\text{fit}}(z)$ in Eq.~\eqref{eq:H0(z)}, we require the following condition
\begin{equation}
    \phi(z)-\left(1+z\right)\,\phi^{\prime}(z)=\left(1-2\alpha\right)\phi(z)\,,
    \label{eq:diff-eq-F}
\end{equation}
which admits the solution:
\begin{equation}
    \phi\left(z\right)=K\,\left(1+z\right)^{2\alpha}\,.
    \label{eq:solution-F(z)}
\end{equation}
We fixed the initial condition $\phi\left(0\right)=K$ at $z=0$, where $K=1-10^{-7}$ \citep{Hu:2007nk} denotes the deviation from a pure GR scenario ($\phi=1$). 

As a consequence of Eq.~\eqref{eq:effective-H0}, the effective Hubble constant becomes
\begin{equation}
    H_{0}^{\text{eff}}(z)=\frac{H_{0}}{\sqrt{K\left(1-2\alpha\right)}\left(1+z\right)^{\alpha}}\,,
    \label{eq:new-effective-H0(z)}
\end{equation}
which is a decreasing function, as requested to match the values of $H_{0}^{\text{loc}}$ and $H_{0}^{\text{CMB}}$ for $z=0$ and $1100$, respectively. In this regard, we set $\alpha=1.1\times10^{-2}$ and $H_{0}=72.2\,\textrm{km s}^{-1}\,\textrm{Mpc}^{-1}$. Note, in particular, that the value of $\alpha$ is consistent in 1 $\sigma$ with the fitting parameters $\alpha=0.009\pm0.004$ used in the analysis of three redshift bins in \cite{dainottiApJ-H0(z)}.

Finally, by substituting $\phi(z)$ from Eq.~\eqref{eq:solution-F(z)} and $H_{0}^{\text{eff}}(z)$ from Eq.~\eqref{eq:new-effective-H0(z)} into Eq.~\eqref{eq:dV/dz}, we can easily integrate to obtain $g(z)$ and reconstruct analytically the scalar field potential. After long but straightforward calculations, we obtain:
\begin{align}
    \Tilde{V}\left(z\right)\, =\, &\Tilde{V}\left(0\right)+\frac{6\alpha}{1-\alpha}\left\{2\left(2+\alpha\right)\frac{1-\Omega_{m0}}{\Omega_{m0}}\ln{\left(1+z\right)}\right.\nonumber\\
    & + \left.\frac{1+2\alpha}{3}\left[\left(1+z\right)^{3}-1\right]\right\}\,,\label{eq:soluzione-analitica-V(z)}
\end{align}
where we set the integration constant $\Tilde{V}(0)=6\left(1-\Omega_{m0}\right)/\Omega_{m0}$, coming from $V\left[\phi\left(z=0\right)\right]= 2\chi\rho_{\Lambda}$.
After solving relation \eqref{eq:solution-F(z)} for $z=z\left(\phi\right)$, we rewrite the potential as
\begin{align}
    \Tilde{V}\left(\phi\right)\, =\, &\Tilde{V}\left(\phi=K\right)+\frac{6\alpha}{1-\alpha}\left\{\frac{2+\alpha}{\alpha}\frac{1-\Omega_{m0}}{\Omega_{m0}}\ln{\left(\frac{\phi}{K}\right)}\right.\nonumber\\
    & + \left.\frac{1+2\alpha}{3}\left[\left(\frac{\phi}{K}\right)^{\frac{3}{2\alpha}}-1\right]\right\}\,.\label{eq:soluzione-analitica-V(phi)}
\end{align}
Such a procedure can be demonstrated to be consistent with the method outlined in \cite{Nojiri:2009kx}, which adapted to our scalar--tensor reformulation amounts to directly integrate in $\phi$ in the equation $\frac{d V}{d \phi}=R$, once the equality $R=6\dot H+12H^2$ and the expressions for $\phi(z)$ and $H(z)$ in Eqs.~\eqref{eq:solution-F(z)} and \eqref{eq:new-effective-H0(z)} are taken into account. An explicit calculation shows that the two results coincide up to the numerical factor $\frac{1-2\alpha}{1-\alpha}\sim 1$, since $\alpha\sim 10^{-2}$, guaranteeing the consistency of the two approaches. This very small discrepancy is due to the approximation we considered in constructing the analytical model (we disregarded the small term $g(z)$ in the modified Friedmann equation \ref{eq:generalized-Friedmann2-with-factor}). The numerical treatment, which follows in the next section, is clearly consistent to the method in \cite{Nojiri:2009kx} up to the desired order of approximation. In particular, a numerical approach is clearly needed to get the function $N(R)$, being $N$ the e-folding variable introduced in \cite{Nojiri:2009kx}, which is not analytically solvable.

\begin{figure}
    \centering
    \includegraphics[scale=0.25]{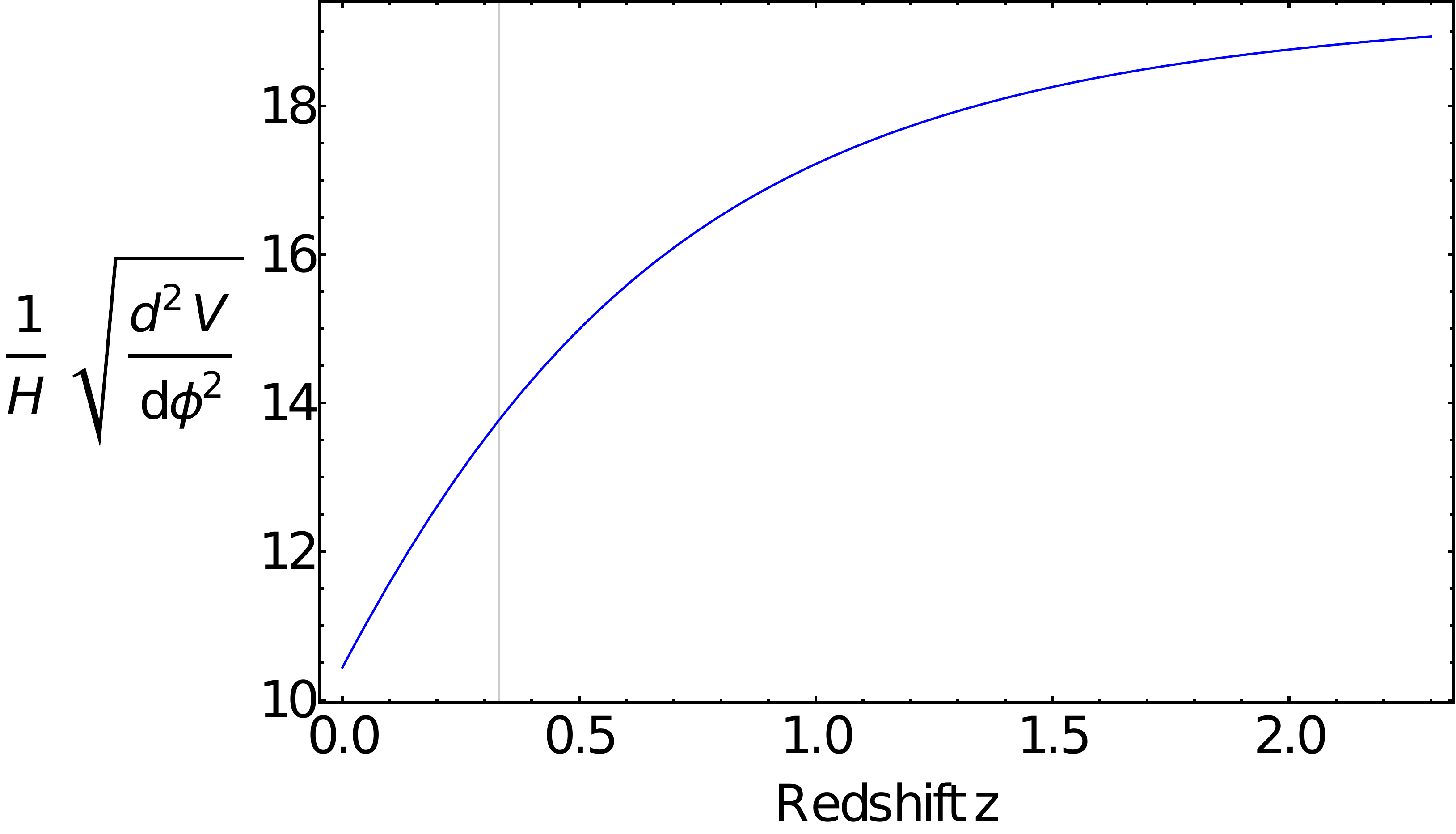}
    \caption{Ratio between the square root of the second derivative of the potential $V=m^{2}\Tilde{V}$ and the Hubble function in terms of the redshift. We have used Eqs.~\eqref{eq:Hubble-parameter(z)},~\eqref{eq:solution-F(z)}, ~\eqref{eq:new-effective-H0(z)}, and ~\eqref{eq:soluzione-analitica-V(phi)}.}
    \label{fig:MsuH}
\end{figure}

The  profile of $\Tilde{V}$ can be appreciated in Fig.~\ref{fig:scalar-field-potential} (we fixed the value $\Omega_{m0}=0.298$ \cite{scolnic-pantheon}), and it can be considered nearly flat for $0<z\lesssim z^*$, where the percentage variation of $\Tilde{V}$ is about $1.6\%$, which validates our hypothesis on a dark energy-dominated era. We conclude this section by noting that in general for $f(R)$ theories the stability of scalar perturbations, i.e. the absence of tachyonic modes in the Jordan frame \citep{Moretti:2019yhs}, implies on a Minkowski background that $\frac{d^2 V}{d \phi^2}>0$ when evaluated in $\phi_{min}$, with $\phi_{min}$ defined by $\frac{dV}{d\phi}=R_{min}=0$. In our case, however, since Eq.~\eqref{eq:soluzione-analitica-V(phi)} is reliable only for a cosmological setting, we can simply look at the behaviour with the redshift of the ratio between the square root of the second potential derivative and the Hubble function, as suggested by \cite{Brax:2008hh}. As illustrated in Fig.~\ref{fig:MsuH}, this ratio is indeed greater than unity for $z=0$, and increases with increasing values of $z$, implying, in agreement with the conclusions of \cite{Brax:2008hh}, that our model is coherent with the requirements of the chameleon mechanism. 

%--------------------------------------------------------------------------------------------------------------
\section{Numerical analysis of the model}\label{sec:numerical}

We now relax the assumption on the existence of a flat region of the scalar field potential for $0<z\lesssim z^*$, but we continue to consider the presence of an effective Hubble constant $H_{0}^{\text{eff}}(z)$.
Let us proceed with a complete numerical analysis of the system \eqref{eq:generalized-Friedmann}~\eqref{eq:scalar-fieldFLRW}, which we want to solve in terms of $\phi(z)$ and $V\left[\phi(z)\right]$.

First, we rewrite the generalized Friedmann equation~\eqref{eq:generalized-Friedmann} in the variable $z$, isolating the dimensionless scalar field potential
\begin{align}
    \Tilde{V}(z)&=6\,\left\{\frac{H_{0}^{\text{eff }2}(z)}{H_{0}^{2}}\left[\left(1+z\right)^{3}+\frac{1-\Omega_{m0}}{\Omega_{m0}}\right]\right.\times\nonumber\\
    &\quad \times \,\left.\left[\phi(z)-\left(1+z\right)\phi^{\prime}(z)\right]-\left(1+z\right)^{3}\right\}\,,\label{eq:Vtilde(z)numerical}
\end{align}
where we used Eq.~\eqref{eq:Hubble-parameter(z)} and the fact that $\rho\sim \left(1+z\right)^3$.

Secondly, we rewrite the scalar field equation~\eqref{eq:scalar-fieldFLRW} as:
\begin{align}
    &\left[\left(1+z\right)^{3}+\frac{1-\Omega_{m0}}{\Omega_{m0}}\right]\left\{\frac{H_{0}^{\text{eff\,2}}(z)}{H_{0}^{2}}\left[\left(1+z\right)^{2}\phi^{\prime\prime}(z)-2\left(1+z\right)\phi^{\prime}(z)\right]\right.\nonumber\\
    & \quad \left.+\frac{H_{0}^{\text{eff}}(z)}{H_{0}^{2}}\frac{d H_{0}^{\text{eff}}(z)}{dz}\left(1+z\right)^{2}\phi^{\prime}(z)\right\}+
    \frac{3}{2}\frac{H_{0}^{\text{eff\,2}}(z)}{H_{0}^{2}}\phi^{\prime}(z)\left(1+z\right)^{4}\nonumber\\
    & \quad -\frac{2}{3}\Tilde{V}\left[\phi(z)\right]+\frac{\phi(z)}{3 \phi^{\prime}(z)}\frac{d\Tilde{V}}{dz}=\left(1+z\right)^{3}\,.\label{eq:scalar-field-eq-numerical}
\end{align}
Then, by substituting $\Tilde{V}(z)$ from Eq.~\eqref{eq:Vtilde(z)numerical} into Eq.~\eqref{eq:scalar-field-eq-numerical} and imposing an effective Hubble constant like in Eq.~\eqref{eq:new-effective-H0(z)}, we obtain a second-order differential equation in $\phi(z)$. We solve numerically this equation with the following initial conditions for $z=0$: $\phi(0)=K$, and $d\phi/dz\,(0)=2\alpha K$. We fixed the same values for $\alpha$, $\Omega_{m0}$, and $K$ adopted in Sect.~\ref{sec:approx}. 

\begin{figure}
    \centering
    \includegraphics[scale=0.25]{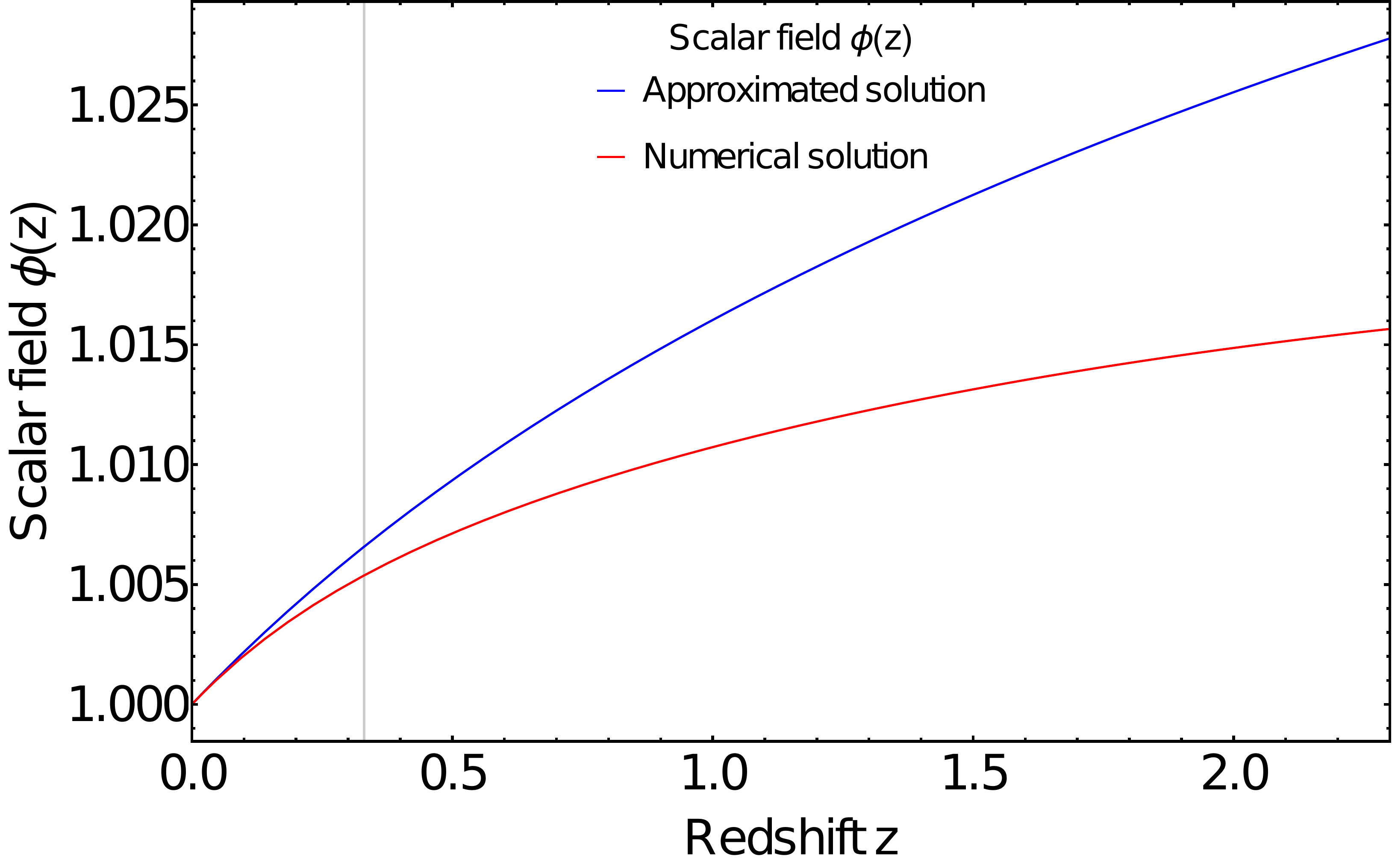}
    \caption{Behaviour of the scalar field $\phi$ versus redshift $z$ in the Jordan frame, assuming an effective Hubble constant $H_{0}^{\text{eff}}\left(z\right)$ in Eq.~\eqref{eq:new-effective-H0(z)}. The blue line is referred to the approximated solution developed in Sect.~\ref{sec:approx}, while the red line is obtained from the numerical analysis discussed in Sect.~\ref{sec:numerical}, after solving Eqs.~\eqref{eq:Vtilde(z)numerical} and \eqref{eq:scalar-field-eq-numerical}. The grey vertical line denotes $z=z^*$.}
    \label{fig:scalar-field}
\end{figure}

\begin{figure}
    \centering \includegraphics[scale=0.25]{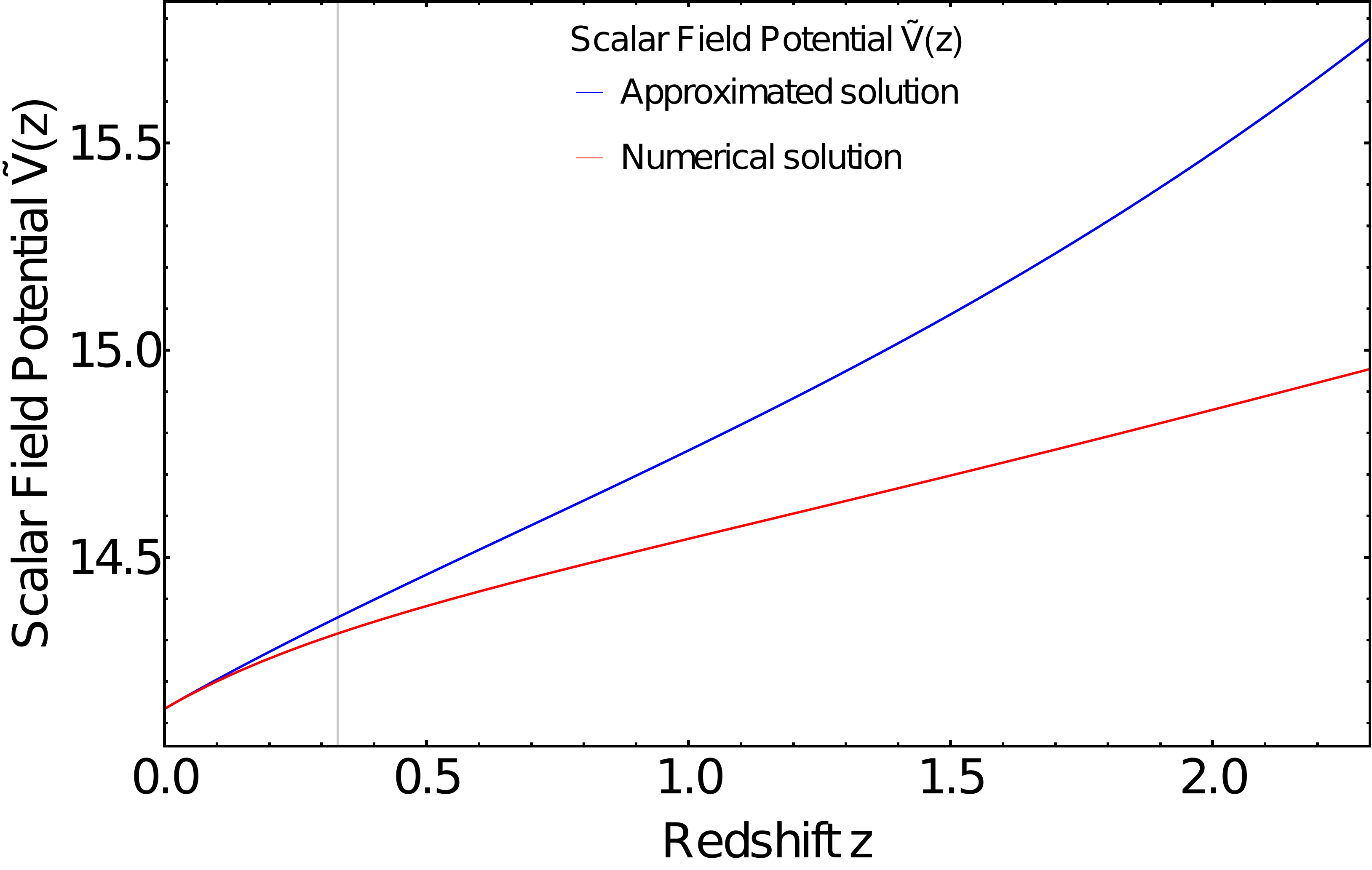}
    \centering \qquad \includegraphics[scale=0.25]{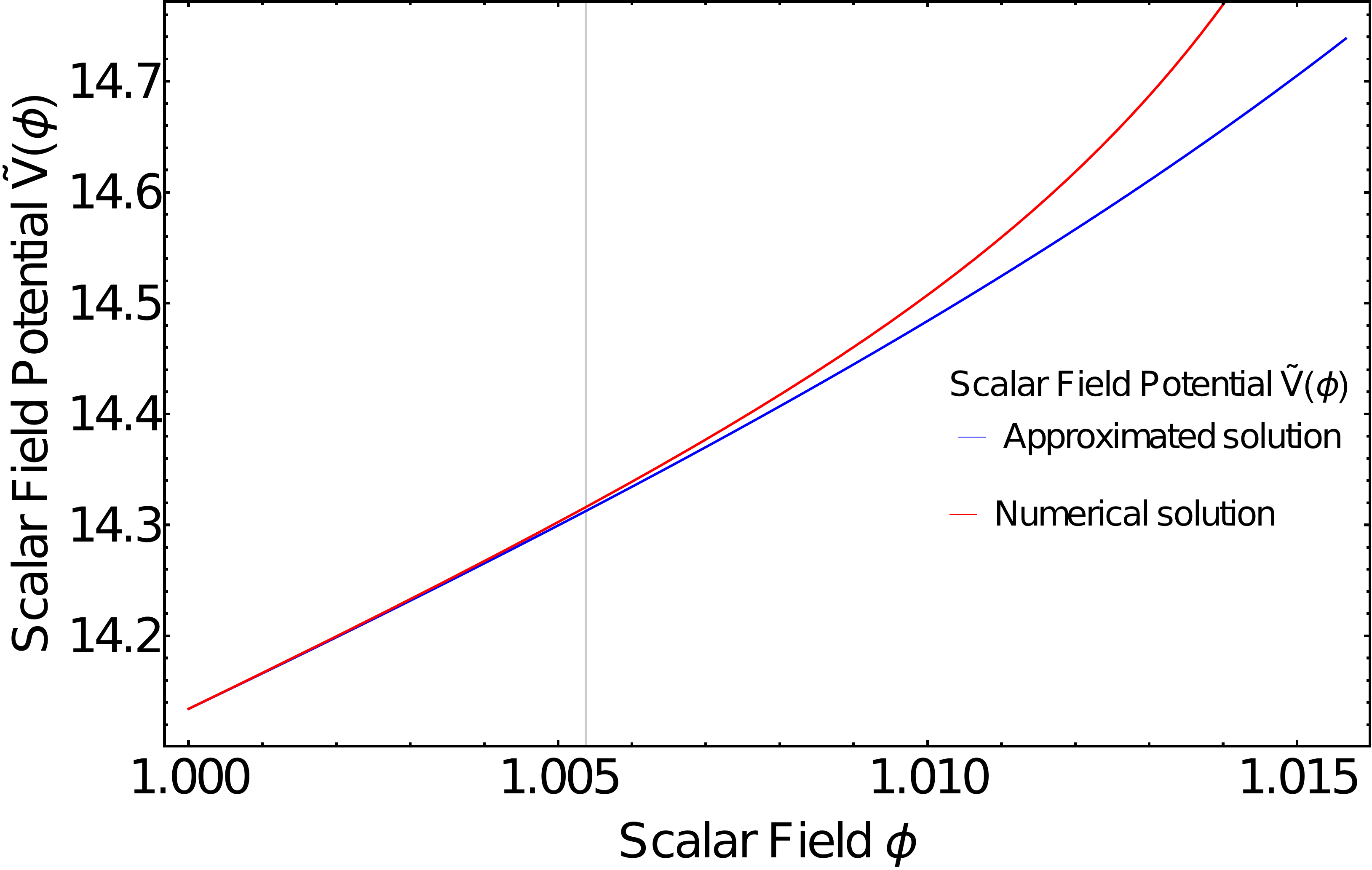}
    \caption{Profile of the scalar field potential in terms of the redshift $z$ (top panel) and the scalar field $\phi$ (bottom panel) in the Jordan frame, inferred from the assumption of a running Hubble constant $H_{0}^{\text{eff}}\left(z\right)$, according to Eq.~\eqref{eq:new-effective-H0(z)}. Note that $\Tilde{V}=V(\phi)/m^2$ is a dimensionless potential. The blue and red lines are referred to the approximated solution (Sect.~\ref{sec:approx}) and numerical results (Sect.~\ref{sec:numerical}), respectively. The grey vertical lines denote $z=z^*$ in the top panel and $\phi=\phi(z^*)$ in the bottom one.}
    \label{fig:scalar-field-potential}
\end{figure}

In Fig.~\ref{fig:scalar-field} we show the evolution of $\phi$ with $z$ using a red line, while in Fig.~\ref{fig:scalar-field-potential} we plot the profile of $\Tilde{V}$ in terms of $z$ and $\phi$. In all these figures, we also compare our numerical results with the respective profiles obtained from the analytical solution based on the assumption of a flat potential at low redshifts in Sect.~\ref{sec:approx}, noting that corresponding solutions mostly overlap for $z\ll1$. It should be stressed that the potential $\Tilde{V}$ exhibits a nearly flat profile for $0<z\lesssim z^*$ also for the numerical solution with a percentage variation of about $1.3\%$.

\section{The low-redshift f(R) profile}\label{sec:lowz}

We are interested in obtaining an analytical expression for the $f(R)$ function, reproducing both the late-time cosmic acceleration and a running Hubble constant with the redshift, according to Eq.~\eqref{eq:H0(z)}.  To this end, we expand the solution for $\phi(z)$ and $\Tilde{V}\left[\phi(z)\right]$ in the limit of low redshifts for $z\ll1$. 

More specifically, starting from Eq.~\eqref{eq:solution-F(z)} for $z\ll1$, we get up to the second order:
\begin{equation}
    \phi(z)\approx K\left[1+2\alpha z-\alpha\left(1-2\alpha\right)\,z^{2}\right]+O\left(z^{3}\right).
    \label{eq:phi(z)lowz}
\end{equation}
{Note that the low-redshift limit $z\ll1$ is equivalent to an expansion for $\phi$ around $K$. Then, we expand $\Tilde{V}\left[\phi(z)\right]$ given by Eq.~\eqref{eq:soluzione-analitica-V(phi)} for $\phi\approx K$:}
\begin{equation}
    \Tilde{V}(\phi)\approx \Tilde{V}(K)+A_{1}\,\left(\phi-K\right)+A_{2}\,\left(\phi-K\right)^{2}+O\left[\left(\phi-K\right)^{3}\right]\,,
    \label{eq:V(phi)lowz}
\end{equation}
where the dimensionless constants $A_{1}$ and $A_{2}$ are defined as
\begin{align}
& A_{1}=\frac{6}{K\left(1-\alpha\right)}\,\left[\frac{1+2\alpha}{2}+\left(\alpha+2\right)\,\frac{1-\Omega_{m0}}{\Omega_{m0}}\right]\,,\label{eq:defA1}\\
& A_{2}=\frac{3}{K^{2}\,\left(1-\alpha\right)}\,\left[\frac{1+2\alpha}{2}\,\left(\frac{3}{2\alpha}-1\right)-\left(\alpha+2\right)\,\frac{1-\Omega_{m0}}{\Omega_{m0}}\right]\,.\label{eq:defA2}
\end{align}
Once we have the expression for $\Tilde{V}\left(\phi\right)$, we use the equation $R=dV/d\phi$ for solving in $\phi=\phi\left(R\right)$. Then, using the relation $f\left(R\right)=R\,\phi\left(R\right)-V\left[\phi\left(R\right)\right]$, we obtain
\begin{equation}
    f\left(R\right)\approx m^2 B_{0}+B_{1}\,R+B_{2}\,\frac{R^2}{m^2}\,,
    \label{eq:f(R)lowz}
\end{equation}
where we have defined the constants
\begin{equation}
    B_{0}=\frac{A_{1}^{2}}{4A_2}-\Tilde{V}\left(K\right)\,, \qquad B_{1}=K-\frac{A_1}{2A_2}\,, \qquad B_{2}=\frac{1}{4A_2}\,.
\end{equation}
It should be stressed that Eq.~\eqref{eq:f(R)lowz} provides an approximated solution of the $f\left(R\right)$ function for $z\ll1$, which contains constant, linear, and quadratic terms in $R$, with the $\Lambda$CDM model recovered for $K\rightarrow1$ and $\alpha\rightarrow0$. Clearly, the function $f(R)$ has been constructed on a cosmological setting and its parameters are not directly suitable for a comparison in the Solar system framework. None the less, the absence of a tachyonic mode, as ensured by the positive coefficient in front of the $R^2$ term, is a reliable consistency check for the theory. 

%--------------------------------------------------------------------------------------------------------------
\section{Conclusions}\label{sec:conclusions}

We started our analysis from the results obtained by \cite{dainottiApJ-H0(z),DainottiGalaxies-H0(z)}, which outlined a dependence of the value of $H_0$ with the redshift via a binned data analysis of the SNe Ia Pantheon sample within 2~$\sigma$. The specific form of the decaying $H_{0}^{\text{fit}}(z)$ given in Eq.~\eqref{eq:H0(z)} was the phenomenological input of our theoretical study. 

The idea proposed above consists in setting up a dark energy model that is able to account for a variation with $z$ of the $H_0$ value. More specifically, we adopted the theoretical paradigm of $f(R)$ gravity, as viewed in the Jordan frame (Sect.~\ref{sec:f(R)inJF}), where we used the non-minimally coupled scalar field for describing the variation of the effective Einstein constant. 

Starting from the equations of motion for an isotropic Universe, we assumed the scalar field as a function of the redshift, and we determined the behaviour of Eq.~\eqref{eq:solution-F(z)} by imposing the desired decaying of $H_0^{\text{eff}}(z)$. Then, by means of the scalar field dynamics, we were able to recover the corresponding potential term, which fixed in turn the $f(R)$ model.
\\ \indent The investigation was performed both analytically and numerically: in the former case, in Sect.~\ref{sec:approx} we assumed the existence of a flat region of the scalar field potential, approximated by a constant value, and then we explicitly determined the potential derivative in Eq.~\eqref{eq:dV/dz}; in the latter case, the scheme was implemented directly on the two basic equations \eqref{eq:generalized-Friedmann} and \eqref{eq:scalar-fieldFLRW}, without any assumption on the potential form. It was rather remarkable that, in both analyses, the potential term singled out a nearly flat region for $z\lesssim0.3$ (Fig.~\ref{fig:scalar-field-potential}), which is exactly when the dark energy contribution of the Universe dominates on the matter content. 

The low-redshift limit of our model in Sect.~\ref{sec:lowz} allowed an analytical determination of the potential term, and hence of the underlying $f(R)$ model. The resulting expression~\eqref{eq:f(R)lowz} for the modified Lagrangian contains a cosmological constant, as well as linear and quadratic contributions in the Ricci scalar. In particular, this result is consistent with other $f\left(R\right)$ gravity models proposed to describe deviations from GR in the $\Lambda$CDM cosmological scenario, without introducing dark energy \citep{STAROBINSKYquadratic,Sotiriou-Faraoni2010,fanizza-f(R),fanizza-quadratic}. 

It is very remarkable for the robustness of our model that this modified scheme approaches the $\Lambda$CDM scenario only when $\alpha \rightarrow 0$ and $df/dR\rightarrow 1$. In other words, even if we reduce the function $H_0^{\text{eff}}(z)$ to a fixed constant value, our model can still contain a small deviation from the $\Lambda CDM$ Universe.

Thus, we can claim that our study is able to simultaneously address two key points: on one hand, we get a modified gravity model as a suitable dark energy candidate; on the other hand, we provided a natural interpretation for the profile of $H_0^{\text{fit}}(z)$ obtained in \cite{dainottiApJ-H0(z),DainottiGalaxies-H0(z)}.

The present study calls attention to further investigations as the redshift increases towards the CMB observations, in order to understand if it can satisfactorily solve the Hubble tension.

\section*{Acknowledgements}
The work of TS was supported by the Della Riccia foundation grant for the year 2023. The work of FB was supported by the postdoctoral grant CIAPOS/2021/169.
% The Acknowledgements section is not numbered. Here you can thank helpful
% colleagues, acknowledge funding agencies, telescopes and facilities used etc.

%%%%%%%%%%%%%%%%%%%%%%%%%%%%%%%%%%%%%%%%%%%%%%%%%%
\section*{Data Availability}
No new data were generated or analysed in support of this research.
 
% The inclusion of a Data Availability Statement is a requirement for articles published in MNRAS. Data Availability Statements provide a standardised format for readers to understand the availability of data underlying the research results described in the article. The statement may refer to original data generated in the course of the study or to third-party data analysed in the article. The statement should describe and provide means of access, where possible, by linking to the data or providing the required accession numbers for the relevant databases or DOIs.



\bibliographystyle{mnras}
\bibliography{paper} % if your bibtex file is called example.bib

%\begin{thebibliography}{99}
%\end{thebibliography}

%%%%%%%%%%%%%%%%%%%%%%%%%%%%%%%%%%%%%%%%%%%%%%%%%%

%%%%%%%%%%%%%%%%% APPENDICES %%%%%%%%%%%%%%%%%%%%%

% \appendix

% \section{Some extra material}

% If you want to present additional material which would interrupt the flow of the main paper,
% it can be placed in an Appendix which appears after the list of references.

%%%%%%%%%%%%%%%%%%%%%%%%%%%%%%%%%%%%%%%%%%%%%%%%%%

% Don't change these lines
\bsp	% typesetting comment
\label{lastpage}
\end{document}